\documentclass[prb,twocolumn]{revtex4-1}

\usepackage{amsmath}  
\usepackage{graphicx}
\usepackage{amsfonts}

\newcommand{\ket}[1]{\left| #1 \right\rangle}
\newcommand{\bra}[1]{\left\langle #1 \right|}

\newcommand{\expect}[1]{\left\langle#1\right\rangle}
\newcommand{\eea}{\end{eqnarray}}
\newcommand{\bea}{\begin{eqnarray}}
\newcommand{\ee}{\end{equation}}
\newcommand{\be}{\begin{equation}}
\newcommand{\half}{\frac{1}{2}}
\begin{document}
\title{Angular momentum in the fractional quantum Hall effect}

\author{S.J. van Enk}

\affiliation{Department of Physics and
Oregon Center for Optical, Molecular \& Quantum Sciences\\
University of Oregon, Eugene, OR 97403}
\date{\today}
\begin{abstract}
Suppose a classical electron is confined to move in the $xy$ plane under the influence of a constant magnetic field in the positive $z$ direction.
It then traverses a circular orbit with a fixed positive angular momentum $L_z$ with respect to the center of its orbit. It is an underappreciated fact that the quantum wave functions of electrons in the ground state (the so-called lowest Landau level) have an azimuthal dependence $\propto \exp(-im\phi) $ with $m\geq 0$, seemingly in contradiction with  the classical electron  having positive angular momentum. We show here that the gauge-independent meaning of that quantum number $m$ is not angular momentum, but that it quantizes the distance of the center of the electron's orbit from the origin, and that the physical angular momentum of the electron is positive and independent of $m$ in the lowest Landau levels.
We note that some textbooks and some of the original literature on the fractional quantum Hall effect do find wave functions that have the seemingly correct azimuthal form $\propto\exp(+im\phi)$ but only on account of changing a sign (e.g., by confusing different conventions) somewhere on the way to that result.
\end{abstract}

\maketitle

\section{Introduction}The integer and fractional quantum Hall effects are well understood and form an established part of 
physics, with the crucial results found experimentally first,\cite{klitzing1980,tsui1982} and described theoretically later.\cite{laughlin1983,jain1990} 

There are not many treatments of the quantum Hall effects at the undergraduate level, but the following three articles are: Ref.~\onlinecite{eisenstein1993}
contains a problem set introducing the basic underlying principles of the quantum Hall effect, Ref.~\onlinecite{johnson2002} provides an explanation of the more complicated fractional quantum Hall effect and the role of wave functions therein, and Ref.~\onlinecite{schober1999} describes an undergraduate experiment on the integer quantum Hall effect.
For excellent lecture notes on the graduate level on both integer and fractional quantum Hall effects, see Refs.~\onlinecite{girvin99,tong2016}. 

The  theoretical description of the Hall effects starts with the quantization of the two spatial degrees of freedom \cite{fn1} of an electron confined to move in a plane (say, the $xy$ plane)
under the influence of a magnetic field perpendicular to that plane (say, pointing in the positive $z$ direction, with magnitude $B$).  
Details of the quantization procedure will be given in the sections following this Introduction. Here, the results are simply stated without derivations (for a textbook description of the more general problem of quantizing charged particles in the presence of a magnetic field and the concomitant issues of gauge invariance, see Ref.~\onlinecite{cohen2006}).

A complete basis of spatial states in two dimensions $\{\ket{n,m}\}$ can be constructed as eigenstates of two commuting hermitian operators, the first of which is the Hamiltonian. These eigenstates are labeled by two corresponding quantum numbers, $n$ and $m$. The first, $n$, has a clear physical meaning: it quantizes the electron's energy in units of $\hbar\omega_B$, with 
\be\label{cyclo}
\omega_B=\frac{eB}{\mu}
\ee
the cyclotron frequency, the frequency with which a classical electron orbits a circle in the $xy$ plane. We use $\mu$ here to denote the electron's (effective) mass so as to avoid confusion with the second quantum number $m$ that is the main focus here. $e>0$ is the elementary charge; the charge of the electron is $-e$.\cite{fn2} 

The (non-relativistic) single-electron Hamiltonian has the form
\be\label{H}
H=\frac{(\vec{p}+e\vec{A})^2}{2\mu},
\ee
which contains the (gauge-dependent) vector potential $\vec{A}$, so that issues of gauge will inevitably show up.\cite{cohen2006}
The eigenenergies $E_n=(n+\half)\hbar\omega_B$
can be found, unsurprisingly, without having to choose a gauge. The states with the lowest energy, those with $n=0$, are known as the lowest Landau levels (abbreviated as LLL).\cite{landau}

The second quantum number, $m$, labels the eigenvalues of a second (independent) hermitian operator, one that commutes with $H$.
In the usual procedure one first fixes the gauge in a nice and convenient way, and  subsequently  chooses an operator that commutes with 
$H$. 
For example, in the so-called Landau gauge $\vec{A}=xB\hat{e}_y$, the Hamiltonian is invariant under translations along the $y$ direction and so the operator $p_y=-i\hbar \partial/\partial y$ commutes with the Hamiltonian. 
Similarly, the choice $\vec{A}=-yB\hat{e}_x$ makes $p_x$ commute with $H$. In the so-called symmetric (or circular) gauge, $\vec{A}=B(-y\hat{e}_x+x\hat{e}_y)/2$, $H$ is rotationally invariant and so
 the $z$ component of the canonical angular momentum, $L_z$,
commutes with $H$. In all three cases the second operator is not gauge invariant, which becomes obvious when one realizes that these three operators do not commute with $H$ for the other two choices of gauge. The corresponding eigenvalues, therefore, do not have a physical, gauge-independent meaning.\cite{cohen2006}
 
For the case of interest here, we consider the symmetric gauge, which is the most convenient gauge choice for describing the fractional quantum Hall effect.\cite{laughlin1983}
In this case one ends up with wave functions whose dependence on the angle $\phi$ [in polar coordinates] is of the form $\exp(-im\phi)$ with $m\geq 0$. This result is a bit surprising given that the electron in the classical limit rotates counter-clockwise (and so has positive angular momentum \cite{note8}), whereas a particle with a wave function  $\exp(-im\phi)$ would have negative or zero angular momentum for $m\geq 0$. 

The solution to this conundrum, as will be shown below in great detail, contains three parts: first, the quantum number $m$ in the usual construction is gauge dependent and has no physical meaning as such. Indeed, wave functions are always gauge dependent, and their gauge degree of freedom consists in a local phase factor. Just as in the Landau gauge a wave function of the form $\exp(iky)$ does {\em not} imply the electron has physical momentum $\hbar k$, so in the symmetric gauge a wave function
$\exp(-im\phi)$ does {\em not} imply the electron has physical angular momentum $-m\hbar$. 

Second, one can calculate what the physical, gauge-independent angular momentum is (relative to the origin) in the states constructed in the symmetric gauge. The expectation value of the $z$ component turns out to be $(2n+1)\hbar$: always positive and independent of $m$.\cite{landau,wakamatsu2018,hitoshi}
Even better, and not very well-known,\cite{kitadono2019} one can construct a gauge-independent operator for angular momentum relative to the center of the circular orbit. We will  find here that the states $\ket{n,m}$ are, in fact, eigenstates of its $z$ component with eigenvalue
$(2n+1)\hbar$, again independent of $m$.

Third, more constructively, one can choose a gauge-independent operator that commutes with the Hamiltonian, and thus construct a set of basis states for which {\em both} quantum numbers have a physical meaning. The physical meaning of the second quantum number found in the symmetric gauge turns out to be that of distance of the center of the orbit to the origin, not angular momentum.

The following two sections repeat in some detail the steps of the standard calculation of the wave functions of the Landau levels in the symmetric gauge. The point is
 to make sure we obtain the correct sign for their azimuthal dependence.
Section \ref{aom} then points out that the sign {\em seems} to be wrong, but then explains why it is, nonetheless, correct.
The physical meaning of
the quantum number $m$ is discussed in Section \ref{gauge}, in which a gauge-invariant set of basis states is constructed. 
Section \ref{sign} identifies different
ways of confusing different sign conventions as they appear in the literature, leading to wave functions of the form $\exp(+im\phi)$ but pertaining to a particle circling in clockwise direction (with, therefore, {\em negative} orbital angular momentum).

\section{Classical Hall effect}
Except for using $\mu$ for the electron's mass, we follow here the notational conventions from Ref.~\onlinecite{tong2016}, those being the most careful notes on the quantum Hall effect the author has found. 

The general solution to the classical equations of motion, $\mu d^2\vec{x}/dt^2=-e d\vec{x}/dt  \times \vec{B}$, with $\vec{B}=B\hat{e}_z$, is 
\bea
x(t)&=&X-R\sin(\omega_B t+\varphi)\nonumber\\
y(t)&=&Y+R\cos(\omega_B t+\varphi)
\eea
with $\omega_B$ the cyclotron frequency, as defined in (\ref{cyclo}).
There are four constants of integration, $R,\varphi,X,Y$.
The energy of an electron in such a circular orbit is
purely kinetic---the potential energy is zero, because the magnetic force does zero work---and its magnitude depends only on one integration constant, $R$:
\be
E^{{\rm clas}}=\frac{1}{2}\mu(\omega_B R)^2.
\ee
This is a highly degenerate class of solutions: the same circular motion (with the same radius $R$ and hence the same energy) can be around any center $(X,Y)$.
This large degeneracy of energy eigenstates shows up in the quantum solution, too.

The classical electron's angular momentum relative to the center of the circular orbit, $(X,Y)$,  points in the positive $z$ direction (or, rather, in the same direction as the magnetic field) and its magnitude is directly proportional to its energy:
\be\label{Lclas}
L_z^{{\rm clas}}=\mu\omega_BR^2=\frac{2E^{{\rm clas}}}{\omega_B}.
\ee
\section{Quantizing the Hall effect}
We describe the magnetic field through the vector potential, i.e., via
$\vec{B}=\vec{\nabla}\times\vec{A}$.
Questions of gauge are inevitable and will become important.

The Hamiltonian of Eq.~(\ref{H}) for a single electron can be rewritten as
\be
H=\frac{\vec{\pi}^2}{2\mu},
\ee
with $\vec{\pi}$ the {\em kinetic momentum} or the {\em mechanical momentum},
equal to $\mu d\vec{x}/dt$ in the Heisenberg picture, and defined as
\be
\vec{\pi}=\vec{p}+e\vec{A}.
\ee
Here $\vec{p}$ is the {\em canonical momentum}.
It's the latter that satisfies (after quantizing the theory)  the standard commutation relations with the position operator. That is, in the position representation we have the usual
$\vec{p}=-i \hbar \vec{\nabla}$.

Under a gauge transformation
\be
\vec{A}\mapsto \vec{A}+\vec{\nabla}\lambda(\vec{r}),
\ee
the wave function of our electron will change, too:
\be
\psi(\vec{r})\mapsto \exp\left(-\frac{ie\lambda(\vec{r})}{\hbar}\right)\psi(\vec{r}).
\ee
An example of a gauge-invariant (and measurable/physical) quantity is
\be
\bra{\psi}\vec{\pi}\ket{\psi}=\mu\frac{d}{dt}\expect{\vec{x}}.
\ee
On the other hand,
\be
\expect{\vec{p}}=
\bra{\psi}\vec{p}\ket{\psi}\mapsto
\bra{\psi}\vec{p}\ket{\psi}-e\vec{\nabla}\lambda(\vec{r})
\ee
is gauge dependent and not measurable.
Now note that the $x$ and $y$ components of $\vec{\pi}$ do not commute with each other:
\bea
[\pi_x,\pi_y]&=&[-i\hbar \nabla_x+eA_x,-i\hbar \nabla_y+eA_y]\nonumber\\
&=&-i\hbar e \nabla_x A_y +i\hbar e\nabla_y A_x=-i\hbar eB.
\eea
Their (gauge-invariant) commutator is in fact a constant (more precisely, a constant times the identity operator), just like the commutator for $x$ and $p$. And since the Hamiltonian $H$ is a sum of the squares of $\pi_x$ and of $\pi_y$ what we get is mathematically equivalent to a 1D simple harmonic oscillator (SHO).
In standard fashion
we can define ``ladder'' operators
\be
a=\frac{1}{\sqrt{2\hbar e B}}(\pi_x-i\pi_y)
\ee
and its hermitian conjugate
\be
a^+=\frac{1}{\sqrt{2\hbar e B}}(\pi_x+i\pi_y)
\ee
such that $[a,a^+]=1$, and so they behave exactly as the lowering and raising operators for the 1D SHO.
Indeed, we can rewrite $H$ as
\be
H=\hbar\omega_B (a^+a+\frac{1}{2}),
\ee
with eigenenergies $E_n=(n+\frac12)\hbar\omega_B$ for nonnegative integer $n$.
These energy levels are referred to as {\em Landau levels}. The $n=0$ states form the lowest Landau level (LLL). 
 We can define a {\em magnetic length}
\be
l_B=\sqrt{\frac{\hbar}{eB}}
\ee
and rewrite $a$
\be
a=\frac{l_B}{\sqrt{2}\hbar} (\pi_x-i\pi_y)
\ee
such that it is
manifestly dimensionless.

We can find a second quantum number $m$ and have a complete set of basis states $\ket{n,m}$ describing the spatial degrees of freedom of our electron, by finding a second hermitian operator that commutes with $H$. Let us first follow the standard procedure and fix the gauge before constructing that second operator. In the symmetric gauge, $\vec{A}=B(-y\hat{e}_x+x\hat{e}_y)/2$, we define a pseudo momentum  as
\be
\tilde{\vec{\pi}}=\vec{p}-e\vec{A}.
\ee
The commutator
\be
[\tilde{\pi}_x,\tilde{\pi}_y]=+i\hbar eB
\ee
differs in sign from $
[\pi_x,\pi_y]$ but is still constant.
We can, therefore, construct another pair of lowering and raising operators
(note the sign difference compared to the definitions of $a$ and $a^+$) by defining
\be
b=\frac{l_B}{\sqrt{2}\hbar}(\tilde{\pi}_x+i\tilde{\pi}_y),
\ee
and its hermitian conjugate
\be
b^+=\frac{l_B}{\sqrt{2}\hbar}(\tilde{\pi}_x-i\tilde{\pi}_y),
\ee
such that $[b,b^+]=1$.
Importantly, the new operators {\em in the symmetric gauge} commute with $\vec{\pi}$. And so we can construct states that are eigenstates of both $H$ (with energy $E_n$) and of $b^+b$ (with nonnegative integer eigenvalues $m$).
Namely, starting with the ``ground state'' $\ket{0,0}$ that is annihilated by both $a$ and $b$ (see Eq.~(\ref{aboo}) below), we construct
\be
\ket{n,m}=\frac{(a^+)^n(b^+)^m}{\sqrt{n!m!}}\ket{0,0}.
\ee
There's a countably infinite degeneracy of each eigenenergy $E_n$.
 We can use the operators $a$ and $b$ to write down explicit wave functions [which inevitably are gauge dependent] that go with the eigenstates $\ket{n,m}$. 
In particular, the state $\ket{0,0}$ satisfies two equations
\be\label{aboo}
a\ket{0,0}=0;\,\,\,b\ket{0,0}=0.
\ee
 We can turn these two equations into differential equations, as follows. Substituting the definitions of $\pi_x$ and $\pi_y$ we have
\bea
a&=&\frac{l_B}{\sqrt{2}\hbar} (\pi_x-i\pi_y)\nonumber\\
&=&\frac{l_B}{\sqrt{2}\hbar}
\left(\frac{\hbar}{i}\left(\frac{\partial}{\partial x}-i\frac{\partial}{\partial y}\right)+\frac{\hbar}{2l_B^2}(-y-ix)\right).
\eea
It is useful to define the two independent complex variables $z=x-iy$ and $\bar{z}=x+iy$ and rewrite
\bea
a&=&-i\sqrt{2}\left(l_B\frac{\partial}{\partial \bar{z}}+\frac{z}{4l_B}\right),\nonumber\\
a^+&=&-i\sqrt{2}\left(l_B\frac{\partial}{\partial z}-\frac{\bar{z}}{4l_B}\right),\nonumber\\
b&=&-i\sqrt{2}\left(l_B\frac{\partial}{\partial z}+\frac{\bar{z}}{4l_B}\right),\nonumber\\
b^+&=&-i\sqrt{2}\left(l_B\frac{\partial}{\partial \bar{z}}-\frac{z}{4l_B}\right).
\eea
Note $b,b^+$ commute with $a,a^+$.
We now obtain the general form of the wave functions for the LLL ($n=0$) states. Namely, $a\psi_0(z,\bar{z})=0$ is solved by 
\be
\psi_0(z,\bar{z})=f(z)\exp(-z\bar{z}/4l_B^2)
\ee
for any analytic function $f(z)$.
The state $\ket{0,0}$ also satisfies
$b\ket{0,0}=0$ and that equation has a similar solution, but with a general analytic function $g(\bar{z})$ as prefactor. The only function that is independent of both $z$ and of $\bar{z}$ is the constant function, and so we conclude the wave function of the $\ket{0,0}$ state has the (properly normalized, Gaussian) form
\be
\psi_{0,0}(z,\bar{z})=\frac{1}{\sqrt{2\pi l_B^2}}\exp(-z\bar{z}/4l_B^2).
\ee
The higher-order wave functions in the LLL can be obtained by applying the raising operator $b^+$ $m$ times, which gives $m$ factors of $z$:
\be\label{w}
\psi_{0,m}(z,\bar{z})=\frac{i^m}{\sqrt{2\pi l_B^2 m!}}
\left(\frac{z}{\sqrt{2}l_B}\right)^m\exp(-z\bar{z}/4l_B^2).
\ee
To conclude this Section we note that the operators $\tilde{\vec{\pi}}$ do not commute with the Hamiltonian in every gauge. This demonstrates that the pseudo momentum is not gauge invariant. And so the physical significance of the eigenvalues $m$ is not clear (yet).
\section{Angular momentum}\label{aom}
The wave functions of Eq.~(\ref{w}) have one peculiar property: in polar coordinates $(\rho,\phi)$ we have $z=x-iy=\rho[\cos\phi-i\sin\phi]=\rho \exp(-i\phi)$, and so
\be
\psi_{0,m}\propto \exp(-im\phi).
\ee
But the electrons classically only move in the counter-clockwise direction and have (on average) positive angular momentum. Shouldn't their wave functions have azimuthal dependence $\propto\exp(+im\phi)$?

First note that wave functions are not gauge independent, and recall that the gauge freedom consists exactly in being able to multiply wave functions by local phase factors. That is, a phase factor $\exp(-im\phi)$ by itself has no physical meaning.
The wave functions are eigenstates of the canonical angular momentum operator $L_z$, which in polar coordinates representation is given by $L_z=-i\hbar \partial/\partial \phi$. But, again, $\expect{L_z}=-m\hbar$ is not gauge invariant and has {\em a priori} no physical meaning. 

Second, we may note that by applying $a^+$ (which {\em is} gauge-invariant) we do get higher-level Landau states with higher ``angular momentum'' in the right direction (we get extra factors $\propto \bar{z}\propto \exp(+i\phi)$.) We may also note that in order to have a complete set of square-integrable 2D wave functions  we do need functions with azimutal dependence $\exp(ik\phi)$ for {\em all} $k\in Z\!\!\!Z$.\cite{fnSB} 

Third if we want to know what angular momentum an electron in the state $\ket{n,m}$ possesses, we should use the {\em kinetic} angular momentum $\vec{r}\times\vec{\pi}$ which is gauge invariant (for more on gauge and angular momentum in this context, see Ref.~\onlinecite{wakamatsu2018}).
In polar coordinates $(\rho,\phi)$ the vector potential in the symmetric gauge is
\be
\vec{A}=\frac{\rho B}{2}\vec{e}_\phi.
\ee
This will give an additional contribution to the kinetic angular momentum proportional to $\vec{r}\times \vec{e}_{\phi}$ which does point in the {\em positive} $z$ direction and  compensates for the negative $-m\hbar$ term one gets from $L_z$.

Let us consider in more detail the $z$ component of the gauge-invariant kinetic angular momentum {\em relative to the origin}
\be
{\cal L}_z=x\pi_y-y\pi_x.
\ee
Using the same operators $a$ and $b$ we defined before, we have in the symmetric gauge 
[once we know we have a gauge-invariant operator, we can perform calculations of its expectation values and eigenvalues in any gauge we find convenient] 
\be\label{Lsym}
{\cal L}_z=\hbar(2a^+a+1-b^+a^+-ba).
\ee
We thus get the gauge-independent result
\be
\bra{n,m}{\cal L}_z\ket{n,m}=(2n+1)\hbar.
\ee
This value is positive and independent of $m$. Note that the states $\ket{n,m}$ are not eigenstates of ${\cal L}_z$: indeed, the kinetic angular momentum does not commute with $H$. Kinetic angular momentum of the electron is {\em not} conserved, in spite of rotational symmetry.\cite{barnett} What is conserved is the total angular momentum of electrons and EM fields together, which includes a contribution from the combination of the electric field the moving electron generates and the external magnetic field.\cite{barnett}

The expectation value of angular momentum in particular other states can be {\em negative} (recall note \onlinecite{note8}) thanks to the presence of the last two terms in (\ref{Lsym}). For example, if we define coherent states $\ket{\alpha,\beta}$ as eigenstates of $a$ and $b$, respectively, with complex eigenvalues $\alpha$ and $\beta$ we get
\be\label{abL}
\bra{\alpha,\beta}{\cal L}_z\ket{\alpha,\beta}=(2|\alpha|^2+1-\beta^*\alpha^*-\beta\alpha)\hbar,
\ee
and this expectation value can be made negative, for example, by choosing $\beta=K\alpha^*$ with $K$ real and $\alpha\neq 0$, and $K$ sufficiently large:
\be
K>1+\frac{1}{2|\alpha|^2}.
\ee
Note this large negative expectation value holds only momentarily, namely when the complex amplitude $\beta$ equals $K\alpha^*$. 
Since $\alpha$ evolves in time but $\beta$ does not  (since $b$ commutes with $H$), that condition is fulfilled only periodically in time.  The time average of the expectation value (\ref{abL}) is positive, and equal to $(2|\alpha|^2+1)\hbar$.

If we define the angular momentum {\em relative to the center of the orbit $(X,Y)$}, then we obtain [see below for the definitions of the operators $X$ and $Y$; see also Ref.~\onlinecite{kitadono2019} for a discussion of this angular momentum]
\be
\tilde{{\cal L}}_z=(x-X)\pi_y-(y-Y)\pi_x=\frac{\pi_x^2+\pi_y^2}{eB}=\frac{2H}{\omega_B},
\ee
and we see that in quantum mechanics, too, this quantity is directly proportional to the Hamiltonian, exactly as we found for the classical electron orbits (cf.~Eq.~(\ref{Lclas})).
Obviously, {\em this} angular momentum is always strictly positive for any state.
\section{Gauge-invariant basis and the physical meaning of $m$}\label{gauge}
Let us return to
the classical parametrization of the circular orbits.
The coordinates of the center of the circle, $X$ and $Y$, correspond in the quantized theory to these two  operators: 
\bea
X&=&x-\frac{\pi_y}{\mu\omega_B},\nonumber\\
Y&=&y+\frac{\pi_x}{\mu\omega_B}.
\eea
Both operators commute with $H$---indeed, $X$ and $Y$ are conserved for the classical orbits---but they do not commute with each other:
\be
[X,Y]=il_B^2.
\ee
Physically, this means we can't quite localize the circular orbits in the $xy$ plane.
Since the operators $X$ and $Y$ are gauge-invariant and commute with the Hamiltonian, we could use them to construct a gauge-independent set of basis states. In fact, since the commutator $[X,Y]$ is a constant, we can define
\bea
c&=&\frac{X+iY}{l_B\sqrt{2}}\nonumber\\
c^+&=&\frac{X-iY}{l_B\sqrt{2}}
\eea
such that $[c,c^+]=1$. Then we can construct eigenstates of the operator $c^+c$ in the usual way, with nonnegative integer eigenvalues, say $k$.
The physical meaning of $k$ is clear, given that
\be\label{meaning}
X^2+Y^2=(2c^+c+1)l_B^2.
\ee
For a state $\ket{n,k}$, the integer $k$ thus tells us how far from the origin the circular orbit is displaced.
This reflects perfectly the degeneracy of the classical solutions. (This result can also be used to count how many states there are in the LLL inside a macroscopic area 
$A$, and one obtains the standard answer ${\cal N}=A/2\pi l_B^2$.)

If we now wish to actually construct wave functions as eigenstates of $c^+c$, we are forced to fix the gauge (recall that wave functions are always gauge dependent). In the symmetric gauge it turns out we actually have $c=-ib$ so that $c^+c=b^+b$. So we simply find exactly the same wave functions as before with $m=k$. And so the gauge-invariant meaning of the quantum number $m=k$ is actually that it quantizes (and quantifies) the distance of the center of the orbit to the origin, as per Eq.~(\ref{meaning}) (see also Ref.~\onlinecite{ballentine1998},  which reaches the same conclusion via a different route).\cite{fnRH}

\section{Conclusions}\label{sign}
Even though the seemingly strange features of the $\exp(-im\phi)$ lowest Landau level wave functions can be explained away, the literature about the fractional quantum Hall effect has remarkably often followed a different path. One change of sign somewhere along the derivation  makes one end up with wave functions (in the symmetric gauge) that behave like $\exp(+im\phi)$.
(The calculations have been done correctly in other parts of the literature,\cite{landau,mikhailov2001} of course, but without comments on the seemingly incorrect sign of the ``angular momentum.'')
Here is a sample of such changes of sign:

Laughlin in his original paper\cite{laughlin1983} chose as single-electron Hamiltonian $H\propto(\vec{p}-e\vec{A})^2$. Now that would be perfectly correct, if only $e$ were negative (or if $H$ were meant as a single-hole Hamiltonian). However, given that he also wrote the (positive!) cyclotron energy explicitly as $\hbar eB/\mu$, $H$ contains the wrong sign, which in turn leads to wave functions $\propto\exp(+im\phi)$. 

That sign in the single-electron Hamiltonian, used with the inconsistent convention of a positive $e$, has been copied many times:
For example, the well-known textbook Ref.~\onlinecite{fradkin2013} states explicitly that the electron charge is $-e$, then states (incorrectly) that the Hamiltonian is $H\propto(\vec{p}-e\vec{A})^2$ and subsequently concludes (incorrectly) that the angular momentum $L_z$ equals a non-negative integer times $\hbar$. The same applies to the textbook \onlinecite{khare2005} on fractional statistics. There are more examples in the literature on the fractional quantum Hall effect (e.g. the seminal Ref.~\onlinecite{jain1990} and also the pedagogical article Ref.~\onlinecite{johnson2002}). 

This particular confusion of sign conventions may well originate from the textbook by Landau and Lifshitz, even though their equations are correct: they give the single-electron Hamiltonian as $\propto (\vec{p}-e\vec{A})^2$ but a footnote several pages later reminds the reader that they use the convention that $e=-|e|$!
 
The lecture notes of Ref.~\onlinecite{girvin99}   perform the correct calculation, but the direction of the magnetic field is flipped just before calculating the wave functions ``in order to get rid of an annoying minus sign.'' Indeed, that leads to the same equations that we derived here, except with $z$ then taking the usual definition of a complex number $z=x+iy$. This leads to wave functions $\propto \exp(+im\phi$), but, of course, by changing the magnetic field direction, the electrons move clockwise and so have negative angular momentum.

The insightful notes on the quantum Hall effect of Ref.~\onlinecite{tong2016} carefully perform the whole calculation correctly, but then in the very end choose as angular momentum operator
$J=i\hbar(x\partial_y-y\partial x)$, correctly concluding that $J\ket{0,m}=m\hbar\ket{0,m}$. $J$, however, equals $-L_z$. Then $z=x-iy$ is incorrectly written as $\rho \exp(+i\phi)$ thus leading to Laughlin's wave functions $\propto \exp(+im\phi)$.

An obvious question is why such (inconsistent)  changes of sign would not be noticed earlier.
The answer probably is that, apart from the incorrect result looking correct, 
(i) the electron's angular momentum in the Hall effect has not been measured (although it has been measured for electron beams propagating through a magnetic field,\cite{schatt2014} and 
(ii) the only $m$-dependent physical property of the lowest Landau levels  one actually makes use of (for example, to count the number of states per unit area)
is the distance of the electron's orbit from the origin. But that distance (rather than angular momentum) happens to be the correct gauge-invariant meaning of the quantum number $m$.
Similarly, for a two-electron Laughlin wave function $\propto (z_1-z_2)^m (z_1+z_2)^M$ neither $m$ nor $M$ is a physical angular momentum (pertaining to the relative and center-of-mass motion, respectively). Rather, $m$ gives the relative distance between one electron and the other's orbit, while $M$ gives the distance of the center-of-mass orbit to the origin.


\begin{thebibliography}{99}
\bibitem{klitzing1980}
K. von Klitzing, G. Dorda, and M. Pepper, ``New method for high-accuracy determination of the fine-structure constant based on quantized Hall resistance," Phys. Rev. Lett. {\bf 45}, 494--497 (1980).

\bibitem{tsui1982} D.C. Tsui, H.L. Stormer, and A.C. Gossard, ``Two-dimensional magnetotransport in the extreme quantum limit," Phys. Rev. Lett. {\bf 48}, 1559--1562 (1982).

\bibitem{laughlin1983}
R.B. Laughlin, ``Anomalous quantum Hall effect: an incompressible quantum fluid with fractionally charged excitations,'' Phys. Rev. Lett. {\bf 50},
1395--1398 (1983).

\bibitem{jain1990}
J.K. Jain, ``Theory of the fractional quantum Hall effect,'' Phys. Rev. B {\bf 41}, 7653--7665 (1990).

\bibitem{eisenstein1993}
J.P. Eisenstein,
  ``The quantum Hall effect,''
 Am. J. Phys., {\bf 61},
179--183 (1993).


\bibitem{johnson2002}
B.L. Johnson, ``Understanding the Laughlin wave function for the fractional quantum Hall effect,''
Am. J. Phys. {\bf 70},  401--405 (2002).



\bibitem{schober1999}
A.M. Schober, B. Ruedlinger, A.J. Dahm,
``Measurement of the ratio h/e 2 in an advanced undergraduate laboratory,''
Am. J. Phys. {\bf 67},
524--527 (1999).

\bibitem{girvin99}
S.M. Girvin,
  ``The Quantum Hall Effect: Novel Excitations and Broken Symmetries,''
arXiv preprint cond-mat/9907002 (1999).


\bibitem{tong2016}
David Tong,
``Lectures on the quantum Hall effect,''
arXiv preprint arXiv:1606.06687 (2016).

\bibitem{fn1}
The electron's spin degree of freedom may be ignored for present purposes.


\bibitem{cohen2006}
C. Cohen-Tannoudji, B. Diu, and F. Laloe, 
\textit{Quantum Mechanics}, p. 315--328 and p. 742--765, Volume I (Wiley, New York, 1977).

\bibitem{fn2}
One point of this paper is to warn of different sign conventions. Landau,\cite{landau} whose name is appended to the energy eigenstates  (consistently!) used the convention  that $e=-|e|<0$ is the charge of the electron. 


\bibitem{landau}
L.D. Landau and E.M. Lifshitz, 
\textit{Quantum mechanics: non-relativistic theory},
(Pergamon Press, Oxford, 1977).

\bibitem{note8}
The classical electron's angular momentum relative to the origin is not conserved and can be negative, but is positive on average. The angular momentum of an electron relative to the center of its orbit is both conserved and positive.

\bibitem{hitoshi}
H. Muruyama, 
``Lectures notes on Landau levels,'' (2006). Available  online at
\url{<http://hitoshi.berkeley.edu/221A/landau.pdf>}.
(2006).

\bibitem{wakamatsu2018}
M. Wakamatsu, Y. Kitadono, and P.M. Zhang,
``The issue of gauge choice in the Landau problem and the physics of canonical and mechanical orbital angular momenta,''
Annals of Physics {\bf 392}, 287--322 (2018).


\bibitem{kitadono2019}
 Y. Kitadono, M. Wakamatsu, and P.M. Zhang,
``Role of guiding centre in Landau level system and mechanical and pseudo orbital angular momenta,''
arXiv preprint arXiv:1905.07569 (2019).

\bibitem{barnett}
C.R. Greenshields, R.L. Stamps, S. Franke-Arnold, and S.M. Barnett,
``Is the angular momentum of an electron conserved in a uniform magnetic field?''
Phys. Rev. Lett. {\bf 113}, 240404 (2014).



\bibitem{mikhailov2001}
S.A. Mikhailov, 
``A new approach to the ground state of quantum Hall systems. Basic principles,''
Physica B: Cond. Matter, {\bf 299}, 6--31 (2001).


\bibitem{fradkin2013}
E. Fradkin, 
\textit{Field theories of condensed matter physics},
(Cambridge University Press, Cambridge, 2013).


\bibitem{khare2005}
A. Khare, 
\textit{Fractional statistics and quantum theory},
(World Scientific, Singapore, 2005).



\bibitem{ballentine1998}
L.E.Ballentine,
\textit{Quantum mechanics: a modern development},
(World Scientific Publishing Company, Singapore, 1998).

\bibitem{fnSB}
Saumya Biswas, private communication.

\bibitem{fnRH}
In the other two familiar (Landau) choices of gauge, mentioned in the Introduction, one finds $p_y=-\mu\omega_B X$ and $p_x=\mu\omega_B Y$, respectively, so that  the corresponding ``momentum-like'' quantum numbers $-k_y$ and $k_x$ specify the $x$ and $y$ coordinates of  the center of the electron's orbit, respectively.

\bibitem{schatt2014}
P. Schattschneider, Th. Schachinger, M. St{\"o}ger-Pollach, S. L{\"o}ffler, A. Steiger-Thirsfeld, K.Y.  Bliokh, and F. Nori, 
``Imaging the dynamics of free-electron Landau states,''
Nature Comm.
 {\bf 5}, 4586-4589 (2014).




\end{thebibliography}
\end{document}